\documentclass[aps,prd,twocolumn,epsf,nofootinbib,superscriptaddress]{revtex4}
\usepackage{graphicx}% Include figure files
\usepackage{amsmath,amssymb,latexsym}
\usepackage{bm}% bold math

\begin{document}

\title{Limitations on the Photo-disintegration Process as a Source of
  VHE Photons}

%\author{Felix Aharonian}
%\affiliation{Max-Planck-Institut f\"ur Kernphysik, 
%             Postfach 103980, D-69029 Heidelberg, GERMANY}

\author{Felix Aharonian}
\affiliation{Dublin Institute for Advanced Studies, 31 Fitzwilliam Place, Dublin 2, Ireland}
\affiliation{Max-Planck-Institut f\"ur Kernphysik, 
             Postfach 103980, D-69029 Heidelberg, GERMANY}

\author{Andrew  M. Taylor}
\affiliation{ISDC, Chemin d'Ecogia 16, Versoix, CH-1290, SWITZERLAND}

\begin{abstract}
\begin{center}

We consider whether photo-disintegration is ever able to provide an effective
mechanism for the production of VHE $\gamma$-ray emission from astrophysical
sources. We find that the efficiency of this process is always smaller by a factor $A/Z^{2}$ 
($\sim 4/A$) than that of nuclei cooling through Bethe-Heitler pair-production.
Furthermore, for sources optically thin to TeV emission, we find that the
efficiency of this process can be no more than $3\times 10^{-5}(R_{\rm source}/R_{\rm Larmor})$,
where $R_{\rm source}$ is the source size and $R_{\rm Larmor}$ is the CR nuclei Larmor radius.
We conclude that this process is unable to provide an effective mechanism
for VHE $\gamma$-ray emission from astrophysical sources.

\end{center}
\end{abstract}

\maketitle

\section{Introduction}

For any candidate cosmic ray (CR) source, the
requirement that the acceleration time in the source be shorter than the
escape time from the source is a condition that must be necessarily satisfied \cite{Hillas2}. 
However, beyond this, other requirements may also be necessarily satisfied
for these sources to also be considered good $\gamma$-ray emitters.
One such condition is that the CR cooling time is not too much longer than 
the CR escape time from the source. 

The possible role of photo-disintegration as an origin of VHE $\gamma$-rays from
astrophysical sources was suggested some time ago \cite{Moskalenko:1987}. 
Further to this, the origin of TeV emission from both star formation regions (SFR) 
\cite{Anchordoqui:2006pe} and gamma-ray bursts (GRB) \cite{Murase:2010va} sources 
have recently been suggested to be produced through photo-disintegration interactions 
of CR nuclei within the source.
Leaving the problem of acceleration aside, 
we here focus instead on the constraints placed by cooling and escape times.
These are considered for photo-disintegration and other available radiation field based 
CR cooling mechanisms in order to determine their relative cooling efficiencies. 

The details of the de-excitation of nuclei following photo-disintegration
interactions \cite{Moskalenko:1989} (such as the exact energy of the de-excitation line
and the photon production multiplicity) 
are of importance for the detailed modelling of the subsequent $\gamma$-ray emission 
spectrum. However, we here consider the cooling times of this process in order to address its relative 
efficiency with respect to other processes in order to determine whether it 
can ever play an important role as a VHE $\gamma$-ray production mechanism.

Since a source's luminosity in secondary particles, $L_{\rm second.}$, 
relates to its CR luminosity, $L_{\rm CR}$, by, 
\begin{eqnarray}
L_{\rm second.} = f L_{\rm CR}\label{luminosities}
,
\end{eqnarray}
where $f=1-e^{-t_{\rm trap}/t_{\rm cool}}$ describes the total cooling
efficiency within the source, $t_{\rm trap}$ are the CR containment times in the source, 
and $t_{\rm cool}$ are the CR cooling times. For sources semi-transparent to CR, 
$t_{\rm trap}/t_{\rm cool}<1$, and $f\approx t_{\rm trap}/t_{\rm cool}$.
For $\gamma$-ray emission powered by such CR cooling, the total efficiency of CR cooling  
relates to the cooling efficiency through the $\gamma$-ray production channel, $f_{\gamma}$, by
$f=f_{\gamma}+f_{e}+f_{\rm other}$, where $f_{\gamma}$ and $f_{e}$ are the cooling efficiencies through
photon and electron/positron production respectively, and $f_{\rm other}$ accounts for the
cooling efficiency through other particle production channels. Though the electron
production channel may not at first glance be considered a radiative channel, these
electrons may proceed to cool in the source's radiation and magnetic fields, giving
rise to radiative emission as they do so.
The specific $\gamma$-ray luminosity relates to the specific CR luminosity by,
\begin{eqnarray}
L_{\gamma}(K_{{\rm CR}\gamma} E_{\rm CR}) = f_{\gamma}L_{\rm CR}(E_{\rm CR})\label{luminosities}
,
\end{eqnarray}
where $K_{{\rm CR}\gamma}$ is the fractional energy exchange of the primary CR to photons
for the relevant process. The factor $f_{\gamma}$ describes the efficiency with which cosmic rays 
are able to transfer their energy directly to photons before they exit the source.

With $\gamma$-ray emission powered by CR in this way, an energy budget crisis may 
be encountered for the parent CR power output if the cooling time is too 
much larger than the escape time. High efficiency in the conversion of CR to $\gamma$-ray 
power is thus frequently required for sources to be considered good $\gamma$-ray emitters. 

The layout of this paper is as follows,
in section~\ref{cooling_mechanism} we consider the CR nuclei cooling efficiency
relative to other radiation field based cooling processes. We follow on from this
by considering in more detail the comparison of these rates with the use of a a fiducial 1~eV 
greybody radiation field. In section~\ref{examples} these results are
applied to typical source environments in order to determine the cooling times of 
CR nuclei and other CR particles for each of these given environments.
In order to see how magnetic fields may improve the source efficiency, 
in section~\ref{magnetic_trapping}, we investigate how far beyond the rectilinear crossing
time ($R_{\rm source}/c$) diffusive CR trapping may increase the escape time by. 
Following this investigation into the efficiency range of these different radiative
processes for particular sources, we investigate the general conditions necessary
for high efficiencies in section~\ref{general}. 
For such maximal efficiencies, the transparency
of our general source to such TeV photon emission through photo-disintegration is determined in 
section~\ref{TeV_opacity}. We finish by reversing these arguments in order to infer a few of the 
necessary source characteristics of UHECR nuclei by requiring that photondisintegration is highly 
inefficient within the source environment.

\section{The Photo-Disintegration Cooling Mechanism}
\label{cooling_mechanism}
Focusing on the CR nuclei cooling efficiency, we here consider
both the photo-disintegration and pair production nuclei cooling channels.
As a comparison of these nuclei cooling processes to other available radiation field
based cooling mechanisms, we also consider CR proton cooling through 
both photo-pion production and pair production channels, and CR electron cooling 
through inverse Compton interactions.

For sources semi-transparent to CR, such that they do not deposit the majority of 
their energy before escaping the source, the (maximum) relative efficiency of a 
given channel, 
$f_{\rm max}^{\gamma/e}$, is dictated by the CR cooling time, $t_{\rm cool}^{\gamma/e}$, 
dependent on the product of the fractional energy exchange to secondary photons or electrons per collision, 
$K_{CR\gamma/e}$ (where the subscript quantities refer to outgoing particles from the collision), 
and the cross-section for these interactions, $\sigma_{CR\gamma}$,
\begin{eqnarray}
f_{\rm max}^{\gamma/e}\propto \left(\left. t_{\rm cool}^{\gamma/e}\right|_{\rm min}\right)^{-1}\approx \left.K_{{\rm CR}\gamma/e}\sigma_{CR\gamma}\right|_{\rm max},
\end{eqnarray}
whose values for each of the processes are given in Appendix~\ref{CR_Int_Cool}.
We note here that cooling processes with threshold energies become highly suppressed
for center-of-mass energies below that required, leading to the possible relative 
enhancement of processes with smaller threshold energies in this case.

For the photo-disintegration process, the fractional energy exchange per interaction 
to photons $K_{A\gamma}\approx \langle N_{\gamma}\Gamma_{A} E_{\star}\rangle /E_{A}$,
where $E_{\star}\sim $1~MeV is the nuclei excitation energy, $N_{\gamma}$ is the
photon multiplicity, and $E_{A}$ is the CR nuclei energy. 
Assuming that a single photon is emitted by the nucleus,
\begin{eqnarray} 
K_{A\gamma}\approx \frac{E_{\star}}{Am_{p}c^{2}}\approx \frac{10^{-3}}{A}.
\end{eqnarray}
Since the photo-disintegration cross-section goes as
$\sigma_{A\gamma}^{\rm th.}\approx A$~mb \cite{Anchordoqui:2006pe} ($\sigma^{\rm th.}$ referring
to the cross-section value in the resonance region just above threshold),
for this process, $\left.K_{A\gamma}\sigma_{A\gamma}\right|_{\rm max}\approx 10^{-30}$~cm$^{2}$,
with a threshold photon energy of $10~$MeV being required in the nucleus' rest frame.
Although upon photo-disintegration nuclei do change species, we here ignore
the effect this has through the assumption that their Lorentz factor is not significantly 
alterered during the disintegration chain.
Interestingly, but for different reasons, a similar fractional energy exchange per interaction
of $K_{Ae} \sim 10^{-3}/A$ is found for nuclei Bethe-Heitler pair-production 
\cite{Bethe-Heitler:book,pair_pair}, 
whose cross-section $\sigma_{A\gamma}\approx Z^{2}$~mb.
This leads to $\left.K_{Ae}\sigma_{A\gamma}\right|_{\rm max}\approx (Z^{2}/A)~10^{-30}$~cm$^{2}$, with a 
threshold photon energy of $1~$MeV being required in the nucleus' rest frame.
Though a photo-meson production channel may also be available for nuclei cooling 
\cite{Michalowski:1977},
we ignore this process in our discussion.

In comparison, for photo-meson production, 
$\left.K_{p\gamma}\sigma_{p\gamma}\right|_{\rm max}\approx 10^{-28}$~cm$^{2}$,
and a photon threshold photon energy of $145~$MeV is required in the proton rest frame. For
proton Bethe-Heitler pair-production, $\left.K_{pe}\sigma_{p\gamma}\right|_{\rm max}\approx 10^{-30}$~cm$^{2}$,
and a photon threshold photon energy of $1~$MeV is required in the proton rest frame.

Similarly, in comparison, for inverse Compton cooling, 
$\left.K_{e\gamma}\sigma_{e\gamma}\right|_{\rm max}\approx 10^{-25}$~cm$^{2}$,
and no photon threshold energy exists. Along with cooling in the source's radiation field, 
CR electrons may also cool via synchrotron cooling in any magnetic fields present. The 
relative rate of electron cooling in these two electromagnetic fields is proportional to 
their energy density
($U_{\gamma}$ and $U_{B}$ respectively). 
In this work, however, we focus our attention only on cooling in the source's radiation fields, 
implicitly assuming that the synchrotron cooling channel in the source environment is subdominant.

From this comparison of photo-disintegration cooling with other cooling
processes available to CR in a radiation field, one at first sight 
would expect that the inverse Compton process is able to provide significantly larger cooling 
efficiency than the other processes. Such a conclusion, however, carries an assumption that this 
cooling channel is free to proceed in the Thomson regime. Furthermore, no strong evidence exists 
that CR electrons, protons, and nuclei are equally populous in the acceleration regions, thus a less
efficient process may well be able dominate the TeV $\gamma$-ray emission from a source if its parent
particle CR population is heavily dominant.
This initial comparison also indicates that pair creation can be expected to 
dominate over the photo-disintegration cooling channel for heavy nuclei. Though this process
can not be thought of as a direct photon emission channel, the high energy electrons
produced may cool either in the same radiation field or in the source's magnetic field. Thus the
electron production channel will give rise to (tertiary) photon emission.

In order to look further into a comparison of these different cooling processes, we next 
determine their radiative cooling rates, explicitly, in a common $\langle \epsilon_{\gamma} \rangle = 1$~eV
greybody environment with a number density of $\langle n_{\gamma} \rangle = 1$~cm$^{-3}$. 
The expressions used to determine these interaction and cooling times are given in 
Appendix~\ref{CR_Int_Cool}.

\begin{figure}[h!]
\centering\leavevmode
\includegraphics[width=0.6\linewidth,angle=-90]{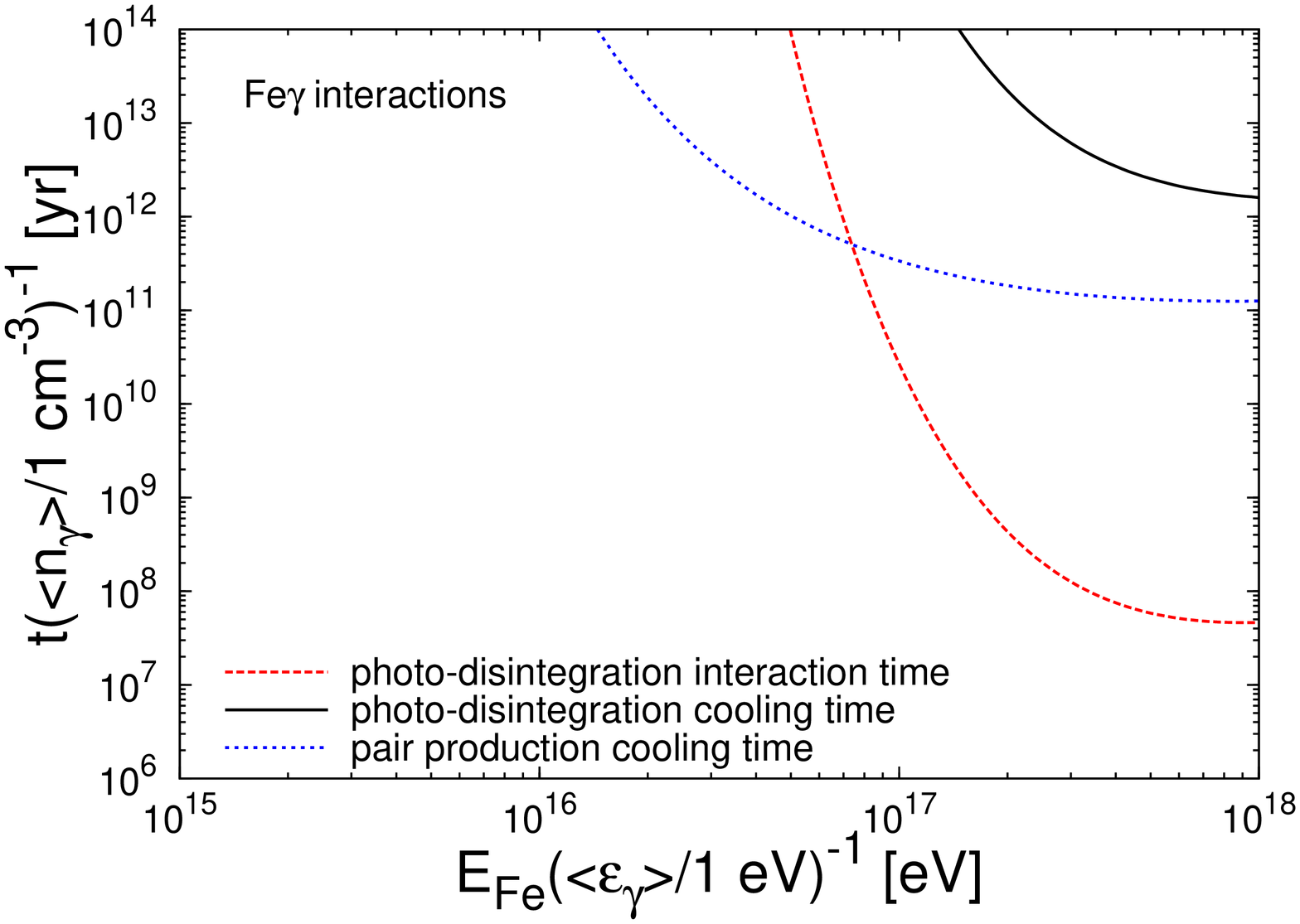}\\
\includegraphics[width=0.6\linewidth,angle=-90]{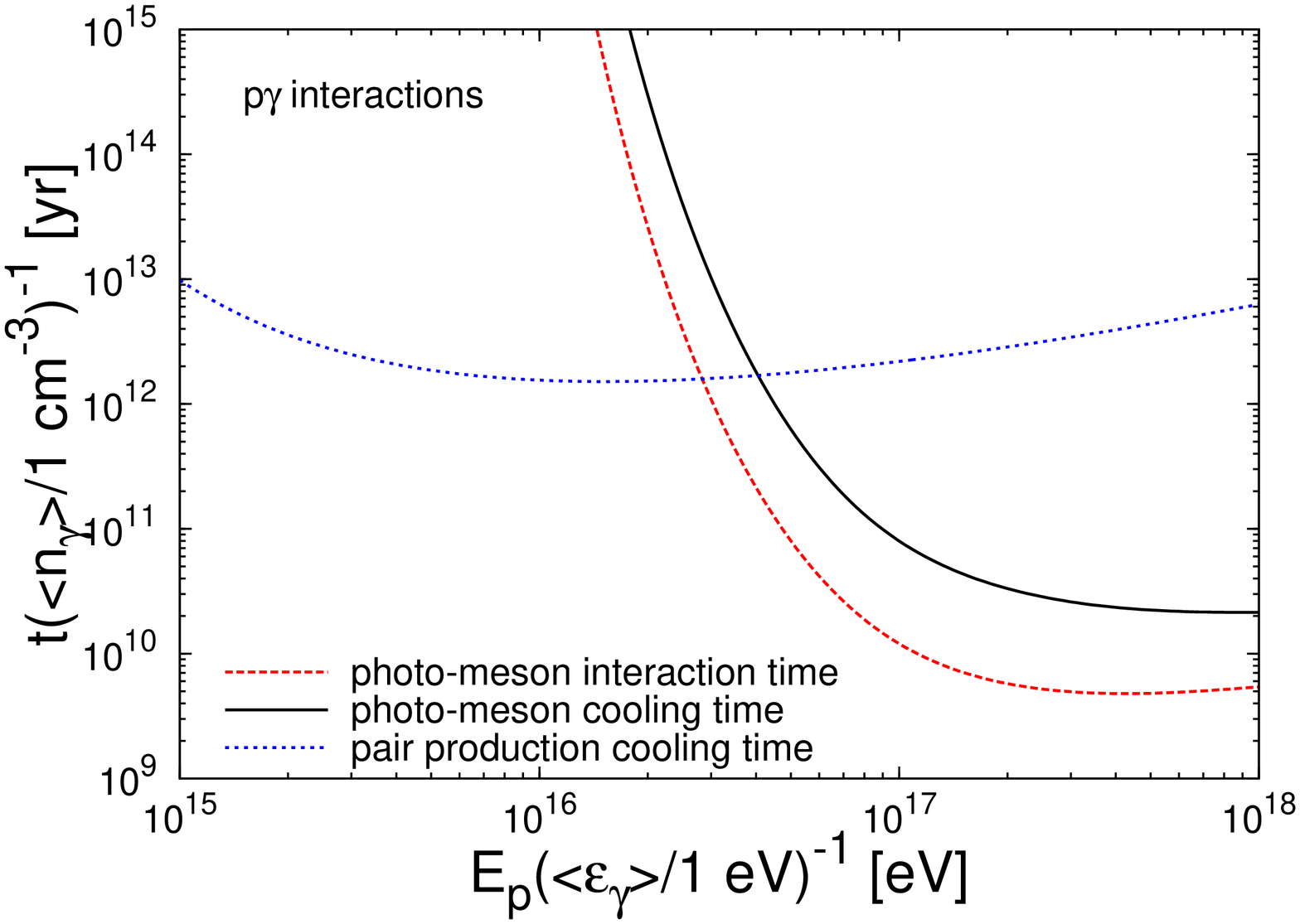}\\
\includegraphics[width=0.6\linewidth,angle=-90]{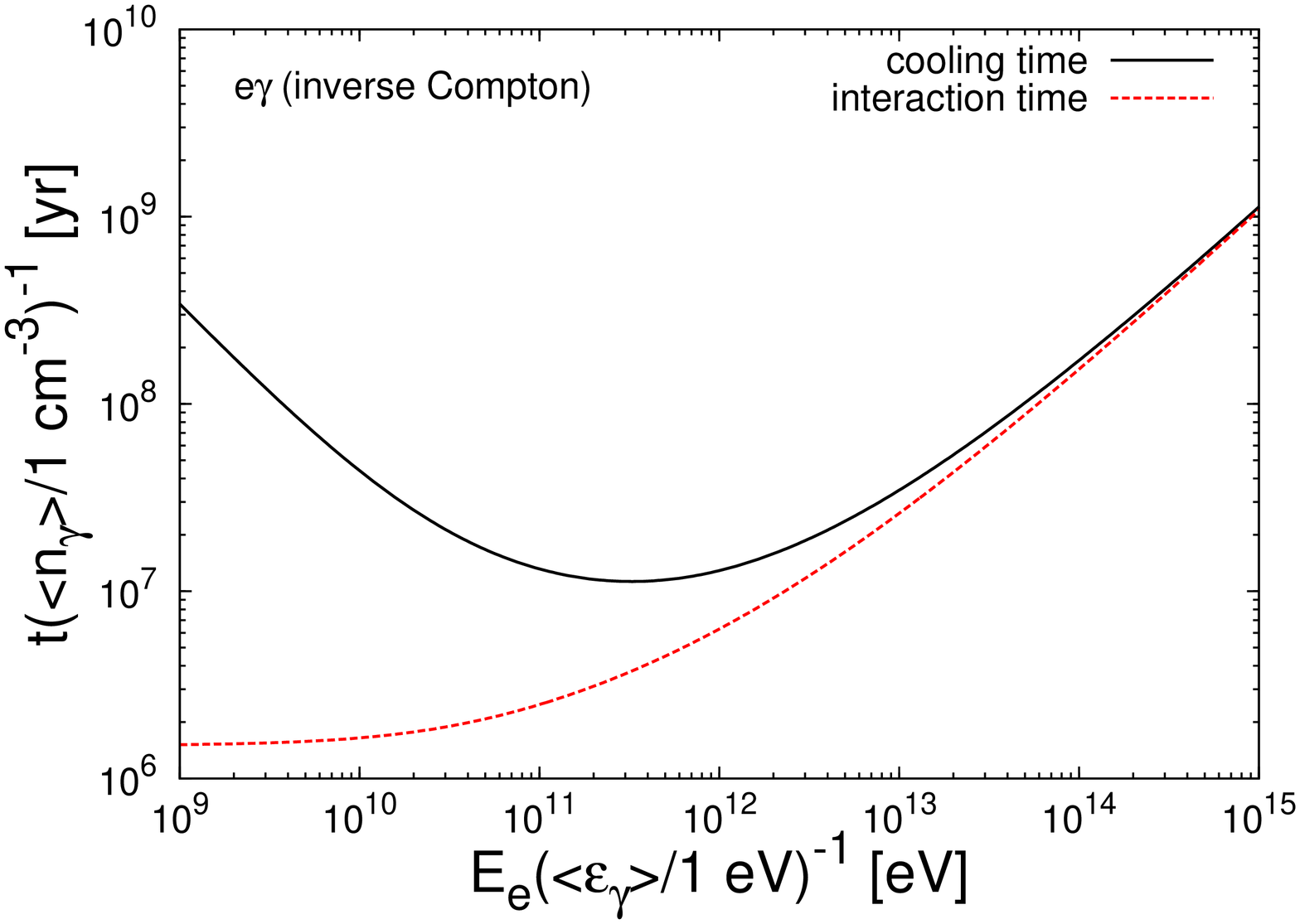}
\caption{The interaction and cooling times for nuclei (top), protons (middle), and electrons (bottom) in our fiducial greybody radiation field with $\langle \epsilon_{\gamma} \rangle=1$~eV and $\langle n_{\gamma} \rangle=1$~cm$^{-3}$.}
\label{times}
\end{figure}

Before a comparison of these cooling curves, with respect to multi-TeV photon
production, can be made, the different energy parent CR must be considered for the different proceses.
With the fractional energy exchange for photo-disintegration being $\sim 10^{-3}/A$,
CR nuclei with an energy of $\sim$500~PeV may give rise to multi-TeV photons through
this process. Similarly, with the fractional energy exchange for photo-production
being $\sim m_{\pi}/m_{p}\approx 0.1$, $\sim$100~TeV protons may also give rise to
multi-TeV photons 
(the ratio between the CR nuclei and protons energies, which give rise to the same
energy photons, being $\sim Am_{p}c^{2}/{\rm MeV }$).
Though the fractional energy exchange to photons for electron IC cooling depends on whether this
occurs in the Thomson or Klein Nishina regime (see Appendix~\ref{CR_Int_Cool}), we consider 
electrons with 100~TeV cooling at the Thomson/Klein Nishina transition, with a fractional
energy exchange of $\sim 10\%$, for the purpose of 
comparison. Thus, in order to for us to fairly compare the multi-TeV emission rates using 
Fig.~\ref{times}, the cooling curve times of $\sim$500~PeV nuclei 
and 100~TeV protons and electrons must be considered.

From the CR nuclei cooling plot in Fig.~\ref{times}, we see that Bethe-Heitler pair production
cooling always dominates over the photo-disintegration channel for heavy nuclei. For
CR proton cooling, however, it is seen that the photo-meson channel dominates over the
pair production channel once this process is able to find sufficient center-of-mass energy 
through collisions with the source photons. 
For electron cooling, a strong minimum is seen to exist in the cooling
time, where, for increasing electron energy, the increase in the fractional energy exchange, 
$K_{e\gamma}$, and the onset in the decrease of the cross-section, $\sigma_{e\gamma}$,
occurs.

\section{TeV $\gamma$-ray Production Efficiencies- Specific Examples}
\label{examples}

We here focus on the efficiencies of the TeV photon production channels available to
the CR nuclei cooling, as well as CR proton and electron cooling, in the radiation field 
of example sources. Although the environments of the sources that we use are not all 
well constrained, we here utilise example values as a demonstrative tool.
Favourable values for all sources are used in these calculations. A brief summary of
these source regions is given in the following paragraph. Further details and assumptions 
about these source regions are given in Appendix~\ref{Source_Environment}.

For the accretion disk (AD) environments surrounding $M\sim$100~$M_{\odot}$ (small) and 
$\sim$10$^{8}$~$M_{\odot}$ (large) blackholes, we assume both their luminosity and 
magnetic fields are at their Eddington values ($L_{\gamma}\approx 10^{38}~(M/M_{\odot})$~erg~s$^{-1}$ 
and $B\approx 6\times10^{8}\eta^{1/2}\left(M/M_{\odot}\right)^{-1/2}~{\rm G}$), with a source
size assumed of $R_{\rm source}\approx 100~R_{\rm Schwarz.}$ in both cases. 
For the gamma-ray bursts (GRB) environment, a bulk Lorentz factor of $\Gamma\approx3\times 10^{2}$ 
and an apparent luminosity of $L_{\gamma}\approx 10^{51}$~erg~s$^{-1}$ are assumed. Along with this, 
an internal (comoving) magnetic field of 
strength 10$^{4}$~G and a (comoving) emission region of size $3\times 10^{-9}$~pc are taken.
For X-ray binary sources (XBS) a luminosity of 10$^{38}$~erg~s$^{-1}$, a
magnetic field of 10$^{2}$~G, and a source region of size $5\times 10^{-6}$~pc is taken. 
Finally, for our star formation region (SFR), a luminosity of 10$^{41}$~erg~s$^{-1}$, a magnetic
field of 10$^{-5}$~G, and a size of 10~pc is used.

Through the application of simple transformations to the resuls in Fig.~\ref{times}, 
using the mean photon density and energy in the source 
(whose values are given in TABLE \ref{sources} of Appendix~\ref{Source_Environment}),
the cooling times for the different source environments are found. 
In this way, Fig.~\ref{times} demonstrates that only the radiation fields of the
more compact sources (such as small AD environments and GRB fireballs) possess 
photons of the necessary energy for all the direct TeV producing channels considered 
to operate. 
For the less compact sources than these (SFR and XRB), with cooler radiation fields, 
only the $A\gamma$ photo-disintegration and $e\gamma$ inverse Compton processes can 
operate as direct TeV $\gamma$-ray producing mechanisms. 

However, assuming that a source's radiation field is sufficiently hot for the direct
photon emission channel to open up, the efficiency of a process depends simply on the
mean photon density $\langle n_{\gamma} \rangle$ of a source's radiation field.
Constraining the minimum CR trapping time in the source , $t_{\rm trap}^{\rm min}$, to be the light 
crossing time for the system, a lower bound can be placed on the source transparency ($f^{{\rm CR}\gamma}$) 
to CR nuclei, protons, and electons for cooling through their direct photon production channel,

\begin{eqnarray}
t_{\rm cool}\approx \left(\frac{4\pi\langle \epsilon_{\gamma}\rangle R_{\rm source}c}{K_{{\rm CR}\gamma}\sigma_{{\rm CR}\gamma}L_{\gamma}}\right)\frac{R_{\rm source}}{c}=f^{{\rm CR}\gamma}_{\rm min}\left(\frac{R_{\rm source}}{c}\right).
\end{eqnarray}

\begin{figure}[h!]
\centering\leavevmode
\includegraphics[width=0.8\linewidth,angle=-90]{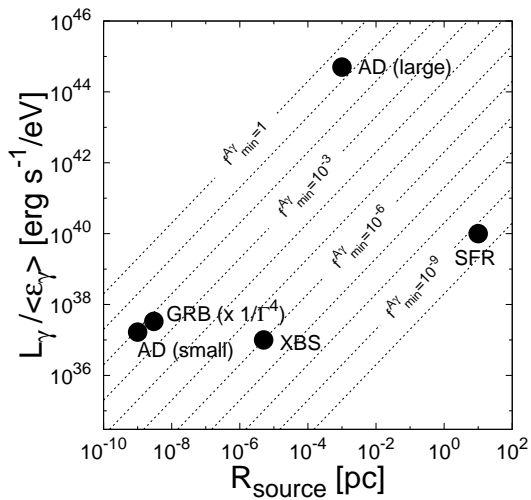}
\caption{A luminosity size diagram showing the source luminosity per mean photon energy in erg~s~$^{-1}$/eV and source size in pc. The dotted lines shown represent lines of constant (minimum) efficiency for the photo-disintegration cooling process. The sources shown are discussed in more detail in Appendix~\ref{Source_Environment}, and are shown as representative examples. 
%This representation reveals the fraction of CR nuclei energy passed on to VHE photons before they leaving the source region under the assumption of straight line propagation though the source.
}
\label{Hillas_new}
\end{figure}

In Fig.~\ref{Hillas_new} we show a comparsion of the example source sizes and 
$L_{\gamma}/\langle \epsilon_{\gamma}\rangle$ values for the photo-disintegration process. 
This plot highlights a problem that both our compact binary and the star formation region 
sources face as CR proton or nuclei sources whose losses in the source's radiation field 
power their TeV photon emission.  For efficiencies, $f_{\gamma}$, far below unity 
(in logarithmic space), larger and larger CR luminosities are necessary to power TeV 
photon emission. With 
$\langle n_{\gamma} \rangle R_{\rm source}$ following a trend of decreasing for larger and larger
source sizes ($R_{\rm source}$), it is clear that a trend of increasingly poor source efficiency 
can be expected, unless the assumption of rectilinear propagation through the source is 
strongly violated. 

In order for the efficiency, $f_{\gamma}$, to be brought closer to unity for the source,
it is necessary that magnetic fields in the source region prolong the duration of CR in the 
source's radiation field (indeed some degree of magnetic field trapping is necessary for 
Fermi first or second order acceleration to take place at all). We consider the maximal 
possible effects magnetic trapping may have on increasing the efficiency in the following 
section.

\subsection{Magnetic Trapping Improvements on the $\gamma$-ray Production Efficiency}
\label{magnetic_trapping}

We constrain the maximum trapping time, $t_{\rm trap}^{\rm max}$, to be the diffusive escape 
time from the system at the Bohm limit , $t_{\rm trap}^{\rm max}=R_{\rm source}^{2}/2D_{\rm Bohm}$, where 
$D_{\rm Bohm}\approx cR_{\rm Larmor}/3$ \cite{Bohm:book}. Thus, an estimation of the factor by which 
magnetic trapping may increase the source efficiency is obtained through a consideration of the 
magnetic fields present in the source regions, by
\begin{eqnarray}
t_{\rm trap}^{\rm max}=\left(\frac{3R_{\rm source}}{2R_{\rm Larmor}}\right)\frac{R_{\rm source}}{c}.
\end{eqnarray}

With CR protons only needing an energy of roughly $(Am_{\pi}c^{2}/{\rm MeV})^{-1}$ times
that of CR nuclei in order to give rise to the same energy photons through photo-meson production 
as those produced through photo-disintegration by CR nuclei of a given energy, their larmor radii
are a factor of $(Am_{\pi}/(Z~{\rm MeV}))^{-1}$ (ie. $\sim 1/200$ when compared to iron nuclei) 
smaller than those of CR nuclei. Thus, for the same source environment, the maximal increase in 
the efficiency of both photo-meson production and inverse-Compton cooling processes are a
factor of $200$ higher than those achievable for nuclei (demanding that all processes
give rise to the same energy TeV $\gamma$-rays).

By presenting, in a ``Hillas plot'' fashion \cite{Hillas2}, the source size and magnetic fields 
for the different source environments we consider (see Appendix~\ref{Source_Environment}), a 
quick determination of how much smaller than the source size the gyroradii of 500~PeV nuclei are 
can be made for each object.

\begin{figure}[h!]
\centering\leavevmode
\includegraphics[width=0.8\linewidth,angle=-90]{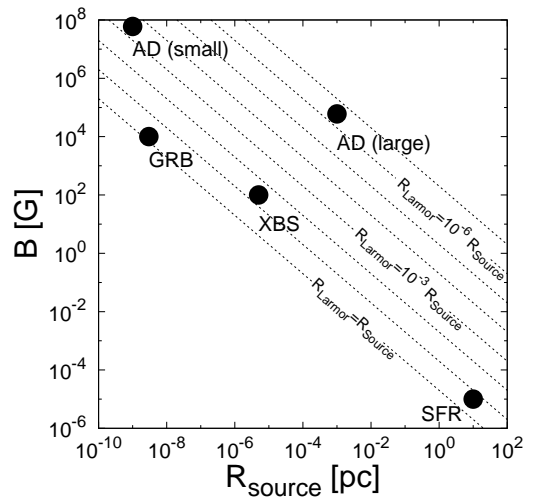}
\caption{A ``Hillas''-type plot \cite{Hillas2} showing the source magnetic field in Gauss and source size in pc. The dotted lines shown represent lines of constant (maximal) trapping enhancement of the efficiency by magnetic fields in the source. The sources shown are discussed in more detail in Appendix~\ref{Source_Environment}, and are shown as representative examples.}
\label{Hillas}
\end{figure}

Using the maximal trapping factors obtained from Fig.~\ref{Hillas}, the increase in the 
photo-disintegration efficiencies for the different sources is obtained. These results
in collaboration with the rectilinear efficiencies found in Fig.~\ref{Hillas_new} allow the
maximum photo-disintegration efficiencies to be obtained.

For the example case of extented sources such as SFR, for which a photo-disintegration origin 
of VHE emission was discussed in \cite{Anchordoqui:2006pe},
a maximal efficiency, taking into account magnetic trapping effects, of $f^{A\gamma}\approx 10^{-8}$ 
is acheived. Such low efficiencies, for objects with TeV luminosities
$\sim 10^{35}$~erg~s$^{-1}$ lead to the requirement of uncomfortably large ($>10^{43}$~erg~s$^{-1}$) 
multi- TeV CR luminosities to power these sources, suggesting that this process is not
responsible for any TeV emission seen from them.
Similary for the example case of compact objects such as GRB, for which a photo-disintegration 
origin of VHE emission was discused in \cite{Murase:2010va}, a maximal efficiency
of $f^{A\gamma}\approx 10^{-2}$ is obtained. Again, a low efficiency for this process 
places a heavy burden on the underlying CR luminosity required.

This does, however, leave open a small window of possibility for a CR nuclei origin of multi-TeV 
photon emission from the more compact GRB and AD sources. 
However, we note two caveats that come with this statement. The first is that even for extragalactic
sources, should photo-disintegration occur with a low efficiency, an uncomfortable demand may
still be placed on the source CR luminosity. Secondly, we also note (as demonstrated
by Fig.~\ref{times}), that photo-distinegration occuring with 
$\left.K_{A\gamma}\sigma_{A\gamma}\right|_{\rm max}\approx 10^{-30}$~cm$^{2}$
and nuclei pair-creation occuring with 
$\left.K_{Ae}\sigma_{A\gamma}\right|_{\rm max}\approx (Z^{2}/A)~10^{-30}$~cm$^{2}$, 
nuclei pair-creation is able to provide the dominant channel for electromagnetic emission, if the 
high energy electrons/positrons are born into sufficiently dense magnetic or radiation fields. 

Thus, the source requirements necessary for maximal efficiencies ($f^{p\gamma}\sim 1$ and 
$f^{A\gamma}\sim 0.1$) for specific cases have been found. In the next section we 
focus on the general source parameters necessary for large efficiencies.

\section{TeV $\gamma$-ray Production Efficiencies- General Case}
\label{general}

We here investigate the conditions required within the environment of a compact luminous object
in order to allow maximal TeV production efficiencies through the photo-meson and photo-disintegration processes.

Using the loss times and the escape times from a system, an
expression for the efficiency is found in terms of the source parameters.
The loss time for CR interactions with background photons in the source may approximately
be expressed as,
$t_{\rm cool}^{{\rm CR}\gamma}\approx 1/(K_{CR\gamma} \sigma_{CR\gamma}^{\rm th.} \langle n_{\gamma}\rangle c)$.
Similarly, the diffusive escape time from the accelerator region is,
$t_{\rm trap}\approx R^{2}/2 D$, where $R$ is the distance being diffused by the CR and
$D$ is the diffusion coefficient.

Thus, using the source efficiency, $f^{{\rm CR}\gamma}\approx t_{\rm trap}/t_{\rm cool}^{{\rm CR}\gamma}$, the photon density may be 
related to the factors describing the diffusive escape and CR interaction physics by,
\begin{eqnarray}
\langle n_{\gamma} \rangle=\frac{f^{{\rm CR}\gamma} 2D}{K_{CR\gamma}\sigma_{CR\gamma}^{\rm th.}R^{2}c}.
\end{eqnarray}

Since, $\langle n_{\gamma}\rangle = L_{\gamma}^{\rm therm.}/c4\pi R^{2}\epsilon_{\gamma}$, the dependence of the efficiency on 
the source luminosity, $L_{\gamma}$, diffusion 
coefficient, $D$, and mean target photon energy $\langle \epsilon_{\gamma}\rangle$ is,

\begin{eqnarray}
f^{{\rm CR}\gamma}=\frac{L_{\gamma}^{\rm therm.}K_{CR\gamma}\sigma_{CR\gamma}^{\rm th.}}{8\pi \langle \epsilon_{\gamma}\rangle D}.\label{luminosity_limit}
\end{eqnarray}

By expressing the diffusion coeffient in terms of the Bohm limit, $D=\xi c R_{\rm Larmor}/3=\xi c E_{CR}/3ZBc$,
\begin{eqnarray}
f^{{\rm CR}\gamma}=\frac{L_{\gamma}^{\rm therm.}K_{CR\gamma}3\sigma_{CR\gamma}^{\rm th.} Z Bc}{\xi 8\pi c\langle \epsilon_{\gamma}\rangle E_{CR}},
\end{eqnarray}
where $\xi$ describes how far the diffusion scattering is from the Bohm limit ($\xi>1$).\\
\newline
{\bf Photo-Meson Production Efficiency}-\\
Since for photo-meson production to occur it is required that,
\begin{eqnarray}
2E_{p}\epsilon_{\gamma}> 2m_{p}c^{2} m_{\pi}c^{2}+(m_{\pi}^{2}c^{2})^{2},
\end{eqnarray}
where $m_{\pi}$ is the rest-mass energy of a pion ($m_{\pi^{0}}c^{2}\sim$135~MeV).
Thus, a simple relationship between the $f^{p\gamma}$, $L_{\gamma}^{\rm therm.}$ and $B$ of a source is found,
\begin{eqnarray}
f^{p\gamma} &\approx &  \frac{ L_{\gamma}^{\rm therm.}cB K_{p\gamma}3\sigma_{p\gamma}^{\rm th.}}{\xi 8\pi c m_{p}c^{2}m_{\pi.}c^{2}}\nonumber\\
% (8 pi \times 938 \times 10^{6} \times 135 \times 10^{6} \times 3\times 10^{10})/(3 \times 5\times 10^{-28})~{\rm eV}^{2}~{\rm s}^{-1}~{\rm cm}^{-1}\nonumber\\
% &=& {7\times 10^{55} ~{\rm eV}^{2}~{\rm s}^{-1}~{\rm cm}^{-1}
 &\approx&  0.3\left(\frac{1}{\xi}\right) \left(\frac{L_{\gamma}^{\rm therm.}}{10^{42}~{\rm erg}~{\rm s}^{-1}}\right)\left(\frac{B}{1~{\rm G}}\right).\label{eff_pgamma}
% erg s^-1 = 0.6\times 10^{12}~eV s^{-1}
% G = 3\times 10^{2}~eV~cm^{-1}
\end{eqnarray}
\newline
{\bf Photo-Disintegration Production Efficiency}-\\
Since for photo-disintegration to occur it is required that,
\begin{eqnarray}
2E_{A}\epsilon_{\gamma}>Am_{p}c^{2} E_{\rm bind.}
\end{eqnarray}
where $E_{\rm bind.}$ is the theshold energy for single nucleon loss ($\sim$10~MeV).
Thus, a simple relationship between the $f^{A\gamma}$, $L_{\gamma}^{\rm therm.}$ and $B$ of a source is found,
\begin{eqnarray}
f^{A\gamma} & = & \frac{L_{\gamma}^{\rm therm.}cB K_{A\gamma}3\sigma_{A\gamma}^{\rm th.}Z}{\xi 4\pi c Am_{p}c^{2}E_{\rm bind.}}\nonumber\\
% (4 pi \times 56\times 938 \times 10^{6}\times 10^{7}\times 3\times 10^{10})/ (3 \times 10^{-27} \times Z)~{\rm eV}^{2}~{\rm s}^{-1}~{\rm cm}^{-1}\nonumber\\
% &=& \frac{2\times 10^{56}}{Z} ~{\rm eV}^{2}~{\rm s}^{-1}~{\rm cm}^{-1}
 &\approx&  0.1\left(\frac{1}{\xi}\right)\left(\frac{Z}{26}\right)\left(\frac{L_{\gamma}^{\rm therm.}}{10^{42}~{\rm erg}~{\rm s}^{-1}}\right)~\left(\frac{B}{1~{\rm G}}\right).\label{eff_Ngamma}
% erg s^-1 = 0.6\times 10^{12}~eV s^{-1}
% G = 3\times 10^{2}~eV~cm^{-1}
\end{eqnarray}

Eqns~(\ref{eff_pgamma}) and (\ref{eff_Ngamma}) succinctly summarise the upper limit results for the
$p\gamma$ photo-meson and $A\gamma$ photo-disintegration TeV producing cooling efficiencies,
agreeing with those obtained from Fig.s~\ref{Hillas_new} and \ref{Hillas} for the different 
sources we considered in the previous section.
However, we note that these expressions do assume that the source radiation field is hot enough
such that these cooling processes are above threshold.
Furthermore, by expressing both a source's luminosity and the magnetic field in terms of their 
Eddington values, for an object of a given mass, 
$L_{\rm Edd.}B_{\rm Edd.}=6\times 10^{46}~(M/M_{\odot})^{1/2}$~erg~s$^{-1}$~G, 
it is seen that the more massive objects offer a larger range of both luminosities and 
magnetic fields that may give rise to large cooling efficiencies through these hadronic
processes. For any such compact massive object with sufficient luminosities and magnetic field 
to achieve large efficiencies, zones will exist around the compact luminous object in which the
radiative efficiency of each of the two processes becomes maximal.

As demonstrated, any region within a source whose luminosity and
magnetic field collectively satisfy the condition $L_{\gamma}^{\rm therm.}B>10^{43}$~erg~s$^{-1}$~G
will have a photo-disintegration channel that is maximally effiecient. 
Furthermore, for such regions the photo-meson channel's efficiency will also
be maximal (and larger).

Thus, a general condition has been found for photo-disintegration to become 
maximally efficient for TeV production mechanism. However, the transparency
of a source to the TeV photons produced from such sources is questionable in
this case. In the following section, we investigate the opacity for maximally 
efficient ($f^{A\gamma}\sim 0.1$) sources to the TeV $\gamma$-rays they produce.

\section{Source Opacity to TeV $\gamma$-rays Produced via Photo-disintegration}
\label{TeV_opacity}

With 500~PeV nuclei needing to interact with photons of energy, $\epsilon_{\gamma}>1$~eV, in order
to give rise to multi-TeV $\gamma$-rays from a source, these same target photons will hinder the escape 
of the TeV photons, through photon photon pair creation interactions if the source is optically
thick to TeV photons. 
In fact, taking the efficiency to be $f^{A\gamma}\approx t_{\rm trap}/t_{\rm cool}^{A\gamma}$, a simple relationship may be 
found relating the radiative efficiency of CR nuclei with energy 
$E_{A}$, to the opacity of the source to TeV $\gamma$-rays, with energy $E_{\gamma}^{\rm TeV}$, 
due to a source radiation field with mean photon energy $\epsilon_{\gamma}$
\begin{eqnarray}
f^{A\gamma} = \left(\frac{R_{\rm source}}{R_{\rm Larmor}(E_{A})}\right)\frac{K_{A\gamma}(s_{A\gamma})\sigma_{A\gamma}(s_{A\gamma})}{\sigma_{\gamma\gamma}(s_{\gamma\gamma})}\tau^{\gamma\gamma},\label{opacities}
\end{eqnarray}
where $\tau^{\gamma\gamma}=R_{\rm source}/ct^{\gamma\gamma}_{\rm cool}$ is the opacity of the source to 10~TeV photons, 
and $s_{A\gamma}=4E_{A}\epsilon_{\gamma}+(Am_{p}c^{2})^{2}$ and $s_{\gamma\gamma}=4E_{\gamma}^{\rm 10~TeV}\epsilon_{\gamma}$ are the squared center-of-mass
energies obtained in the two sets of collisions. For $\epsilon_{\gamma}=1$~eV,
$s_{A\gamma}\approx s_{\rm A\gamma}^{\rm th.}$ and $s_{A\gamma}\approx 40~s_{\gamma\gamma}^{\rm th.}$
respectively, resulting in eqn~(\ref{opacities}) leading to the relation,
\begin{eqnarray}
f^{A\gamma}\approx 3\times 10^{-5} (R_{\rm source}/R_{\rm Larmor})\tau^{\gamma\gamma}.
\end{eqnarray}
For this result we have utilised the relation
$\sigma_{\gamma\gamma}(40~s^{\rm th.})\approx 0.15~\sigma_{\gamma\gamma}^{\rm max}$.
Hence, if $R_{\rm Larmor}>3\times 10^{-5}~R_{\rm source}$,
$\tau^{\gamma\gamma}=1$ is a necessarily satisfied condition for ensuring maximal radiative efficiency.
On the other hand, if $R_{\rm Larmor}<3\times 10^{-5}~R_{\rm source}$, maximal efficiency is still possible
for $\tau^{\gamma\gamma}<1$. 
Thus, for $R_{\rm source}B<1$~pc~G, the attenuation of TeV photons by the source's radiation field
is a natural consequence of a maximal radiation efficiency ($f^{A\gamma}\approx 0.1$). Applied to
the example sources considered in section~\ref{examples}, all but the large AD source would attenuate 
their TeV emission, if this emission originates from efficient photodisintegration interactions.

If the source's radiation field dominates sufficiently over its magnetic field energy
density (taking into account the reduction in $\sigma_{e\gamma}$ at these energies), such attenuation 
will lead to the development of an electromagnetic cascade in the source.
The feeding of these heavy nuclei CR cooling powered cascades, in fact, will be dominanted by the 
injection of electron/positron pairs produced through Beth Heitler nuclei pair production interactions.
Once initiated, these cascades will continue to develope until the cascade photons have insufficient energy
to further photon photon pair create. 
Photons of energy $E_{\gamma}<100 (\epsilon_{\gamma}/1~{\rm eV})^{-1}$~GeV will thus be able to 
eventually escape the source's radiation field. 
Such sources may therefore be expected to exhibit the
universal cascade spectrum shape at these energies \cite{Zdziarski:1988}.
If, however, the source's magnetic field dominates sufficiently over its radiation field energy density,
the energy deposited into the electron/positron pairs will be re-released via sychrotron 
emission with an energy $E_{\gamma}^{\rm sync}\approx 1~(B/1~{\rm G})$~MeV.

Thus, we find that only CR sources more than five orders of magnitude larger than
the Larmor radii of the 500~PeV CR they are accelerating are able to be transparent to
TeV photons and still be efficient to the photo-disintegration cooling process which produced
the TeV photons. All other sources with large photo-disintegration efficiencies will kill their 
own TeV emission either through the feeding of electromagnetic cascades down to 
100$(\epsilon_{\gamma}/1~{\rm eV})^{-1}$~GeV 
energies or through MeV synchrotron emission. This all but closes the window on TeV emission powered
by photo-disintegration, leaving only regions close to a large AD remaining viable source
regions where both the efficiency may be large and the TeV opacity small enough for this
emission to escape.

In our final section, our previous arguments are turned on their head, with us instead
demanding that a source is inefficient to photo-disintegration in order for it to be a
good candidate source of the UHECR nuclei detected at Earth.

\section{The Low Efficiency Requirement for the Sources of UHECR Nuclei Detected at Earth}
\label{UHECR_nuclei}

We here briefly consider the ease with which both UHECR nuclei and and TeV photons may 
propagate through the source radiation field. At these higher CR energies, the UHECR nuclei 
interact preferentially close to threshold with 10$^{-2}$~eV IR photons. Furthermore, any 
multi-TeV photons produced by such interactions will themselves interact close to 
threshold with the same target photons.
Thus, a similarity relation holds between $10^{20}$~eV iron nuclei
photo-disintegration interactions and $10^{14}$~eV $\gamma$-ray pair production interactions,
as discussed in \cite{Neronov:2007mh,Murase:2010va}. 

With the peak cross-section for heavy nuclei, just above threshold, being $\sigma_{A\gamma}^{\rm th.}\sim$A~mb, 
and that for photon photon pair production, just above threshold being $\sigma_{\gamma\gamma}^{\rm max}\sim$200~mb, the 
interaction rates of these two processes go
approximately as $t^{Fe}_{\rm int}\approx 4~t^{\gamma}_{\rm int}$.
Since both processes, close to threshold, occur with photons of a similar energy ($\sim 10^{-2}$~eV), 
the detection of VHE (100~TeV) photons from astrophysical objects,
without evidence of attenuation through pair production in the source, may be used as an indicator that such
a source is also transparent to UHECR nuclei. 

More quantitatively, applying eqn~(\ref{opacities}) for both 
processes close to threshold, giving us the relation,
$f^{A\gamma}\approx 5\times 10^{-6} (R_{\rm source}/R_{\rm Larmor})\tau^{\gamma\gamma},$
combining this with the constraint that if no significant photodisintegration takes 
place in the source, $f^{A\gamma}<10^{-4}$, tells us that for a source transparent to UHECR nuclei,
\begin{eqnarray}
\tau^{\gamma\gamma}< 20 (R_{\rm Larmor}/R_{\rm source}).
\end{eqnarray}
Thus, it is possible for such a source to be be opaque to TeV photons,
and yet still be sufficiently transparent to UHECR nuclei so as not to significantly 
disintegrate them. Depending on how much smaller than the source the Larmor radii of $10^{20}$~eV
nuclei are, it is clearly also possible for a source to be transparent
to TeV photons and not to UHECR nuclei. However, with few astrophysical objects
even able to satisfy the Hillas criterion for $10^{20}$~eV nuclei, it is
hard to envisage $R_{\rm source}/R_{\rm Larmor}$ being considerably larger than a
factor of 10 or so. Thus it seems most natural that an UHECR source transparent
to TeV photons will be transparent to UHECR nuclei.

Furthermore, if TeV photon emission is powered by photo-disintegration, the TeV
luminosity of a source transparent to both UHECR nuclei and TeV photons must be $\sim f^{A\gamma}$
times the sources UHECR nuclei luminsity.
With the local extragalactic 1-100~TeV $\gamma$-ray sources within $\sim 50$~Mpc being
M~87, Cen~A, and NGC~253, M~82, having TeV luminositites $\leq 10^{41}$~erg~s$^{-1}$ \cite{tevcat}, 
their TeV power output approximately match that required by the UHECR sources
if a source number density of $10^{-5}$~Mpc$^{-3}$ is assumed \cite{DeMarco:2005ty}.
Thus, the similarities in these luminosities is strongly suggestive that their 
multi-TeV $\gamma$-ray emission is not powered by UHECR nuclei photo-disintegration.

We lastly note as a matter of interest that if it is assumed that an accelerator of 
UHECR nuclei has its magnetic luminosity in equipartition with its thermal luminosity 
\cite{Waxman:2003uj,Levinson:2009kh}, an assumption perhaps not so obviously motivated, 
the luminosity of 10$^{20}$~eV iron nuclei must be 
$L_{\gamma}^{\rm therm.}\approx 4\times 10^{42}$~erg~s$^{-1}$. Applying to this eqn~(\ref{eff_Ngamma}) 
with $f^{A\gamma}\approx 10^{-4}$ suggests that the acceleration site of 10$^{20}$~eV iron nuclei
must contain magnetic fields of the order of $4\times 10^{-5}$~G in strength!

\section{Conclusion}

We have here followed the suggestion that the photo-disintegration channel, 
available to CR nuclei cooling in a radiation field, may function as an 
effective TeV production mechanism. We have taken a closer look at this
process through the consideration of the cooling times and subsequent
efficiency in converting CR luminosity into $\gamma$-ray luminosity.
A comparison of the cooling rates for CR nuclei relative to those 
for CR proton and electron cooling was made using the
$K_{{\rm CR}\gamma}\sigma_{{\rm CR}\gamma}$ value for each process. 
This strongly indicated that the photo-disintegration channel for CR nuclei
was highly inefficient, being neither larger than those of the other species, nor
larger than that for the Bethe Heitler pair production cooling channel.

Following these comparisons, we determined more precisely the cooling efficiency for 
CR nuclei (and CR protons and electrons) for specific example sources.
Small photo-disintegration efficiencies were found for all but the most compact 
luminous objects. A consideration of the magnetic trapping enhancement on 
these efficiencies indicated that objects for which the efficiency was 
small did not have their efficiency significantly improved by such trapping.
Further to this, the photo-disintegration channel efficiency was found to be always 
a factor $A/Z^{2}$ smaller (10\% for iron nuclei) than that of the Bethe Heitler 
pair production channel efficiency.

The production efficiencies for the general case were then obtained for both
the photo-meson CR proton and photo-disintegration CR nuclei cooling mechanisms.
This result verified that a source for which photo-disintegration operates best
requires both a large luminosity and magnetic field 
($L_{\gamma}^{\rm therm.}B>10^{43}$~erg~s$^{-1}$~G). 
However, the subsequent absorption of the TeV emission by the radiation field in such 
sources was found to be an inevitable problem in almost all cases. Furthermore,
the transparency of a source to its TeV emission was found to put strong
constraints on the photo-disintegration efficiency, with
$f^{A\gamma}<3\times 10^{-5}(R_{\rm source}/R_{\rm Larmor})$ for such sources, 
where $R_{\rm Larmor}$ is the Larmor radius of a 500~PeV nuclei in the source,
which has sufficient energy to give rise to multi-TeV $\gamma$-rays through
this channel.

With low TeV $\gamma$-ray production efficiencies being obtainable through
photo-disintegration, uncomfortable demands are placed on the underlying CR luminosity 
powering the system in almost all scenarios.
Thus, we conclude that this process is unable to offer the efficient conversion of CR 
luminosity into multi-TeV luminosity frequently required by astrophysical sources.

Turning these arguments around for UHECR nuclei sources, the transparency of a 
source to multi-TeV $\gamma$-rays was shown to be a strong indicator that 
a source is also transparent to UHECR nuclei. Furthermore, the comparison of
the CR luminosity expected from nearby extragalactic regions with the multi-TeV
luminosity of nearby extragalactic sources was used to demonstrate that 
the photo-disintegration channel can not be responsible for this
multi-TeV emission.

%\begin{widetext}

\begin{appendix}

\section{Source Size and Environment}
\label{Source_Environment}

{\bf Accretion Disk Fed Radiation Fields}-
We here consider the radiation fields surrounding accretion disks (ADs).
Following the assumption that the radiation fields around these objects
are dominated by their accretion disk's thermal emission, we determine
the expected radiation field for regions close to the ADs
surrounding blackholes. This description is used to obtain the radiation
field for the family of blackhole masses, from
small AD ($\sim 10^{2}M_{\odot}$) to large AD ($\sim 10^{8}M_{\odot}$) environments.

For the small AD environment, we use a luminosity of $10^{40}$~erg~s$^{-1}$.
% R_Schwarz (100 M_sol) = 2.94 x 10^5 m
For a fiducial size of these systems, we consider a region a factor of
10 larger than the AD, whose radius is 
assumed to be 10~$R_{\rm Schwarz.}$, giving us a source size of $\sim 10^{-9}$~pc.
% R_Schwarz (10^8 M_sol) = 2.94 x 10^11 m
Similarly, for the large AD environment we assume a mass of $\sim$10$^{8}$~$M_{\odot}$, an 
AD of 10~$R_{\rm Schwarz.}$, and a source size of $\sim 10^{-3}$~pc.

Assuming that the luminosity of the object is at the Eddington limit,
$L_{\rm Edd.}=10^{40}~(M/100~M_{\odot})$~erg~s$^{-1}$, and that this radiation is released via
an AD at $10~R_{\rm Schwarz.}(100~M_{\odot})$, the blackbody radiation field 
energy density in the vicinity of the AD is,
% R_Schwarz = 2.94 km (M/M_sol)
% L_Edd = 1.26 x 10^38 erg s^-1 (M/M_sol)
\begin{eqnarray}
U_{\gamma} &=& \frac{L_{\rm Edd.}}{4\pi (10~R_{\rm Schwarz.})^{2}c}\left(\frac{R}{10~R_{\rm Schwarz.}}\right)^{-2}\\
 &=& 2\times 10^{23}~\left(\frac{R}{10~R_{\rm Schwarz.}}\right)^{-2}~{\rm eV~cm}^{-3}.
\end{eqnarray}
where $R$ is the radial distance from the star's surface.

If this energy flux is released via blackbody emission from the AD surface, 
the mean photon energy of this radiation will be,
\begin{eqnarray}
\langle \epsilon_{\gamma} \rangle = 600~\left(\frac{100~M_{\odot}}{M}\right)^{1/4}~{\rm eV}
\end{eqnarray}

Thus for small AD we employ a radiation field with $\langle n_{\gamma} \rangle=2\times 10^{18}$~cm$^{-3}$ 
and $\langle \epsilon_{\gamma} \rangle = 600$~eV, and for large the AD
$\langle n_{\gamma} \rangle=1\times 10^{14}$~cm$^{-3}$ and $\langle \epsilon_{\gamma} \rangle = 20$~eV. 
\newline

{\bf Stellar Fed Radiation Fields}-
We also consider objects whose radiation fields are dominated
by stellar emission such as certain high-mass % low or high?
% AU = 1.4 x 10^11 m
X-ray binary sytems (XBS), with a size $\sim$AU ($5\times 10^{-6}$~pc). 
Our fiducial object is assumed to consist of a high mass companion star and compact 
neutron star binary system. 
%For a detailed description of these objects %better word?- alternative to system 
%see \cite{? ask Mitya}. 
The radiation fields present within these systems are dominated by the light emitted 
from the stellar surface of the massive companion star, which we take to be
30~$M_{\odot}$ in mass, 8~$R_{\odot}$ in size, emitting blackbody radiation with a mean energy of 
$\langle \epsilon_{\gamma}\rangle =10$~eV. Thus, the radiation field density in the vicinity of
this star will be,
\begin{eqnarray}
% 1.9 \times 10^{15}~cm^{-3}
\langle n_{\gamma} \rangle = 2\times 10^{15} \left(\frac{\langle \epsilon_{\gamma} \rangle }{10~{\rm eV}}\right)^{3}\left(\frac{R}{8~R_{\odot}}\right)^{-2}.
\end{eqnarray}
where $R$ is the radial distance from the star's surface.
Such stars bathe the system in $\langle \epsilon_{\gamma}\rangle \approx $10~eV energy photons with number 
densities of the order $\langle n_{\gamma} \rangle \approx 10^{11}$~cm$^{-3}$ close to the 
star ($\sim 1000~R_{\odot}$).
\newline

{\bf An Ensemble of Stellar Fed Radiation Fields}-
A third category of object to be considered are star formation regions (SFR),
with sizes $\sim$10~pc. These systems contain even more dilute radiation fields 
than the systems previously considered, consisting of an ensemble
of tens of thousands of stars, of which hundreds are massive stars ($>$30~$M_{\odot}$) 
\cite{Brandner:2008vg}, with a collective luminosity which we estimate to
be $\sim 10^{41}$~erg~s$^{-1}$. 

 % a size of 10~R_{\odot} is used here
The radiation field in these systems is dominated (in energy density) by the 
output from the ensemble of $\sim$1000 massive stars in the region.
As discussed in \cite{Anchordoqui:2006pe}, the dilution factor, $d$, 
of the radiation field from an ensemble of stars of radius $R_{\rm OB} (\approx 10~R_{\odot})$ 
distributed randomly within a sphere, 
goes with the ratio of the collective surface area of the ensemble of stars to
the total surface area of the encompassing sphere surrounding these stars.
% note- need to alter d to balance books
With the system being $4\times 10^{7}R_{\rm OB}$ in size, $d\approx 5\times 10^{-12}$.
Thus, the average energy density, averaged over the whole region, will be $d\times U_{\gamma}$, where
$U_{\gamma}$ is the black body energy density at the stellar surface, which 
for $M=30~M_{\odot}$ with $\langle \epsilon_{\gamma}\rangle =10$~eV, has a value 
$U_{\gamma}\approx 2\times 10^{16}$~eV~cm$^{-3}$, is $d U_{\gamma}\approx 10^{5}$~eV~cm$^{-3}$. 
\newline

{\bf Synchrotron Fed Radiation Fields}-
For our GRB radiation field model, we continue in the vein of \cite{Halzen:2002pg}, 
applying the results of the fireball model, and assuming the GRB radiation field to be the 
result of  synchrotron emission from a relativistic population of electrons. 
For our fiducial object, we take an emission region of size (in the
comoving frame) $\Gamma c \Delta t\approx 3\times 10^{-9}$~pc, using
$\Gamma=3\times 10^{2}$ and $\Delta t=10^{-3}$~s.
Assuming the GRB mean power output to be 10$^{51}$~erg~s$^{-1}$,
with the emission each $\Delta t\approx 10^{-3}$~s being released by a different shell,
and the mean photon energy in this emission being $\langle \epsilon_{\gamma}\rangle \approx 10^{6}$~eV, 
the total number of emitted photons released per shell is
\begin{eqnarray}
N_{\gamma}=\frac{L_{\gamma}\Delta t}{\langle \epsilon_{\gamma}\rangle}\approx 6\times 10^{53}
\end{eqnarray}

Taking the dimensions of the emitting shell in the lab frame to have
width $\Gamma^{2} c \Delta t$ and thickness $c \Delta t$, an estimate of the 
lab frame volume of the emitting region is,
\begin{eqnarray}
V=4\pi \Gamma^{4} \left(c\Delta t\right)^{3}\approx 3\times 10^{33}~{\rm cm}^{3}
\end{eqnarray}

Thus, the estimated lab frame number density of the radiation is $2\times 10^{20}$~cm$^{-3}$. In the
comoving frame of the shell, in which the shell's thickness is $\Gamma R_{\rm Schwarz.}$, 
this reduces to $\langle n_{\gamma} \rangle \approx 7\times 10^{17}$~cm$^{-3}$, and the
mean photon drops to $3\times 10^{3}$~eV. %In comparison
A summary of the radiation field densities and mean
photon energies in each of these sources is given in TABLE~\ref{sources}.

%\begin{widetext}
\begin{table}[H]
\begin{center}
\begin{tabular}{|c|c|c|c|}
\hline
Candidate Object & Size [pc] & $\langle n_{\gamma} \rangle $[cm$^{-3}$] & $\langle \epsilon_{\gamma}\rangle$ [eV] \\
\hline
\hline
GRB & $3 \times 10^{-9}$ & $7\times 10^{17}$ & $3\times 10^{3}$ \\
AD (small) & $10^{-9}$ & $2\times 10^{18}$ & $6\times 10^{2}$ \\
AD (large) & $10^{-3}$ & $10^{14}$ & $20$ \\
XBS & $5 \times 10^{-6}$ & $10^{11}$ & $10$ \\ 
SFR & $10$ & $10^{4}$ & $10$ \\
  \hline
\end{tabular}
\end{center}
\caption{A summary table of the source sizes, radiation field densities, and mean photon energies.}
\label{sources}
\end{table}
%\end{widetext}

\section{CR Interactions and Cooling}
\label{CR_Int_Cool}

{\bf Inverse Compton Scattering Interactions}-\\
The cross-section ($\sigma_{e\gamma}$) for this process is given by \cite{Akhiezer:book},
\begin{eqnarray}
\sigma_{\gamma e}(b)=\frac{3}{4}\frac{\sigma_{\rm T}}{b}\left[\left(1-\frac{4}{b}-\frac{8}{b^{2}}\right)\ln(1+b)+\frac{1}{2}+\frac{8}{b}-\frac{1}{2(1+b)^{2}}\right]\nonumber
\end{eqnarray}
%\end{widetext}
where $b$ is the squared center-of-mass energy available in the collision in units of
$\left(m_{e}c^{2}\right)^{2}$, which for head on collisions is given by 
$b=4E_{e}\epsilon_{\gamma}/(m_{e}c^{2})^{2}$.
This cross-section in the Thomson regime ($b\ll 1$), $\sigma_{\rm T}$, has a
value of 665~mb (=6.65$\times 10^{-25}$~cm$^{2}$), and a fractional energy exchange per
interaction (the fraction of energy the particle gives up to photons per collision), 
which grows with squared center-of-mass energy ($b$) as \cite{Aharonian:1981},
\begin{eqnarray}
K_{e\gamma}(b) = \frac{\left<E_{\gamma}^{\rm out}\right>}{E_{e}} &=& \frac{1}{E_{e}\sigma (b)}\int_{0}^{E_{\gamma}^{\rm out}}E_{\gamma}^{\rm out}\frac{d\sigma_{e\gamma}(b,E_{\gamma}^{\rm out})}{dE_{\gamma}^{\rm out}}dE_{\gamma}^{\rm out}\nonumber\\
&=& f(b)\frac{b}{1+b}
\end{eqnarray}
where $E_{\gamma}^{\rm out}$ is the energy of the photon produced, $E_{e}$ is the energy
of the interacting electron,  
$f(b)$ is an angle averaged factor, taking into account the momentum distribution of
the radiation field, and is of the order of 1/3 in the Thomson limit $(b\ll 1)$.

We note here that we neglect the opening up of other channels for $e\gamma$
interactions, such as triple pair production, since these channels are suppressed
by $\sim \left(\alpha/\pi\right)$, and don't compete with the inverse Compton channel until
energies well above those for which $b=1$.

Inverse Compton interactions can lead to the rapid cooling of VHE electrons and the
subsequent production of VHE photons. 

\begin{figure}[h!]
\centering\leavevmode
\includegraphics[width=0.33\linewidth,angle=-90]{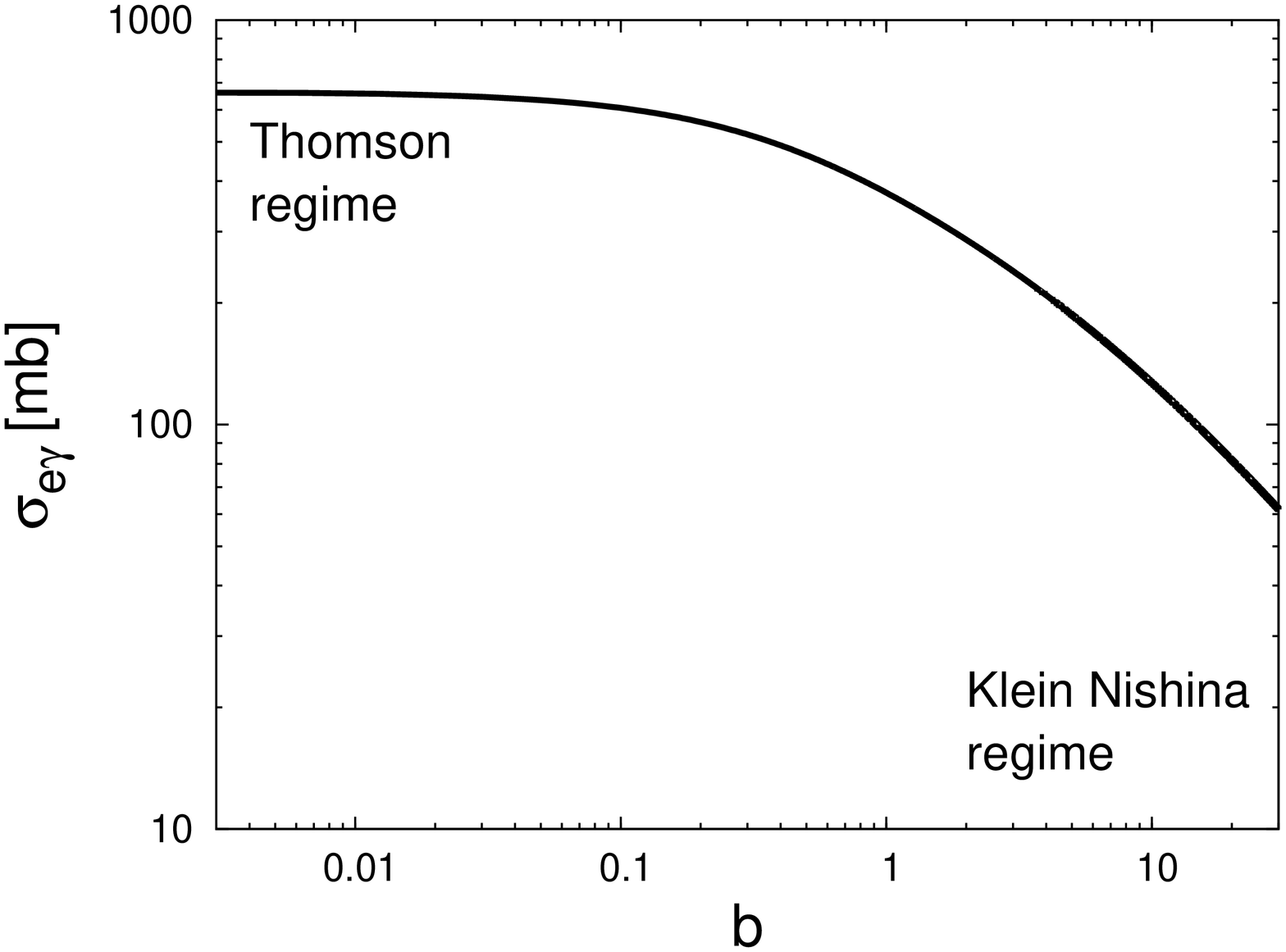}
\includegraphics[width=0.33\linewidth,angle=-90]{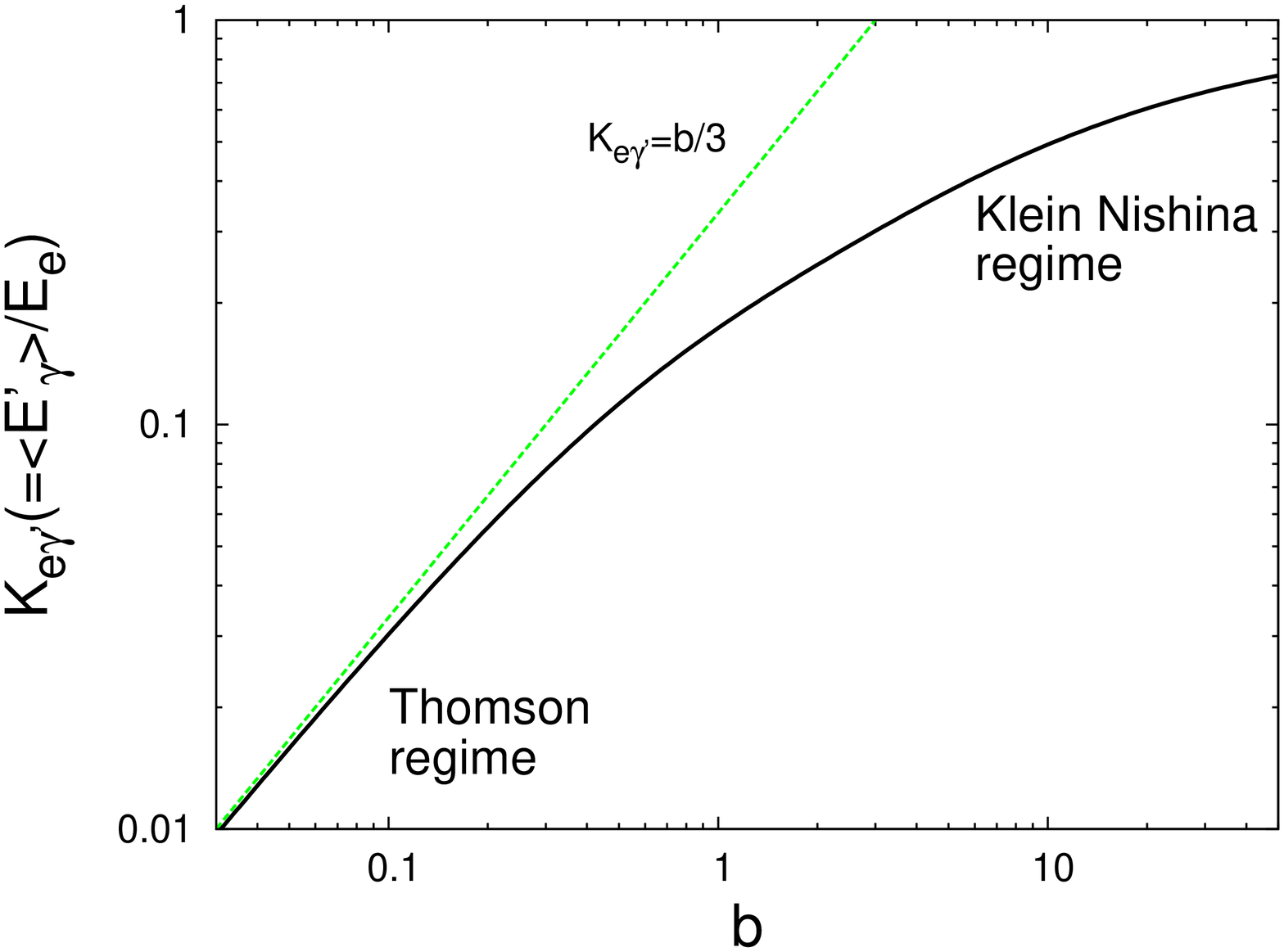}
\caption{The cross-section (left) and fraction energy exchange per interaction (right) for inverse Compton interactions with an isotropic photon distribution as a function of the squared center-of-mass energy in the collision in units of $(m_{e}c^{2})^{2}$, a quantity we here refer to as b.}
\label{IC_inelasticity}
\end{figure}

{\bf Pion Production Interactions}-\\
The cross-section ($\sigma_{p\gamma}$) for this process has been determined using 
SOPHIA \cite{Mucke:1999yb}, whose model fits well the measured data \cite{webpage}.
The energy exhange to photons per interaction has also been determined using SOPHIA.

\begin{figure}[h!]
\centering\leavevmode
\includegraphics[width=0.33\linewidth,angle=-90]{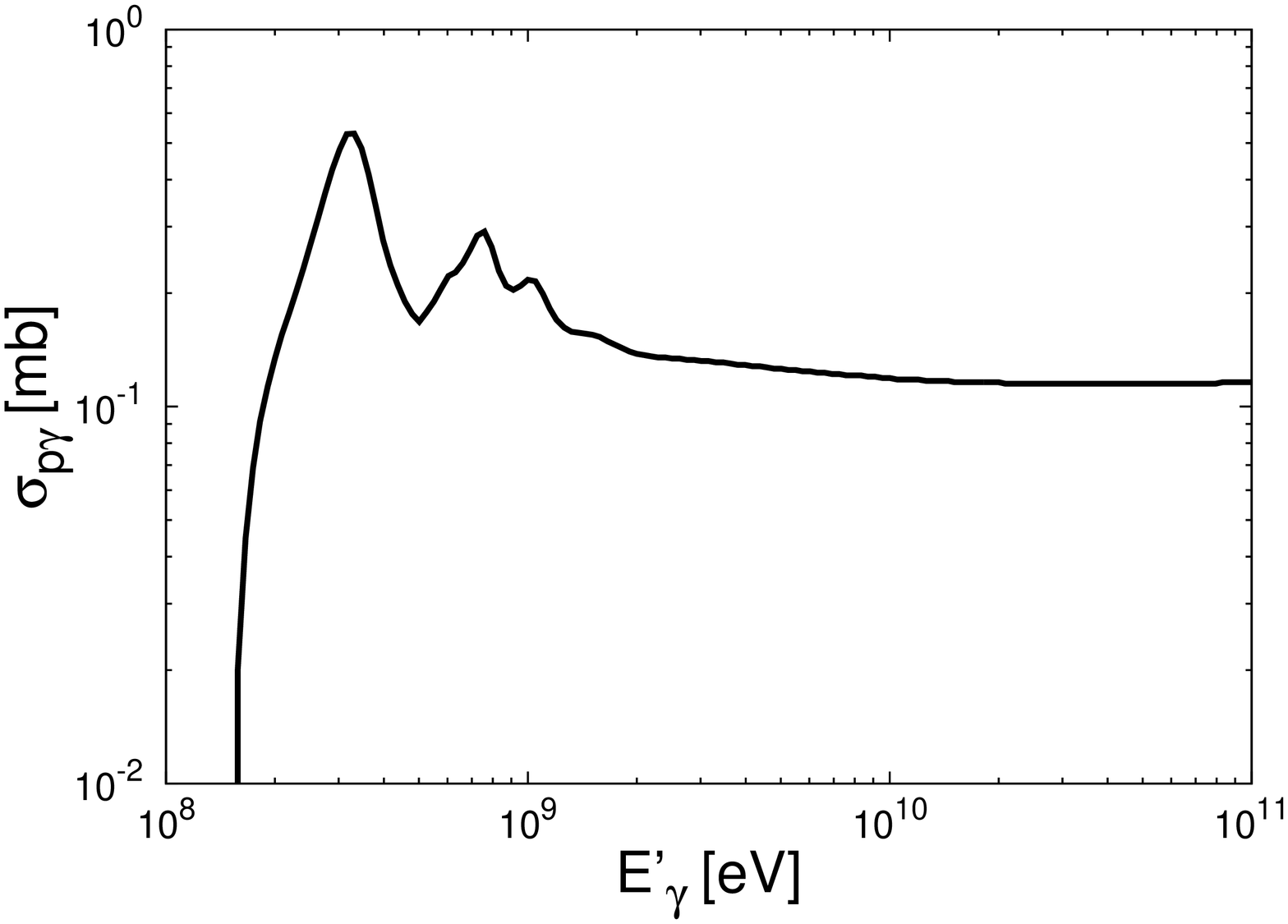}
\includegraphics[width=0.33\linewidth,angle=-90]{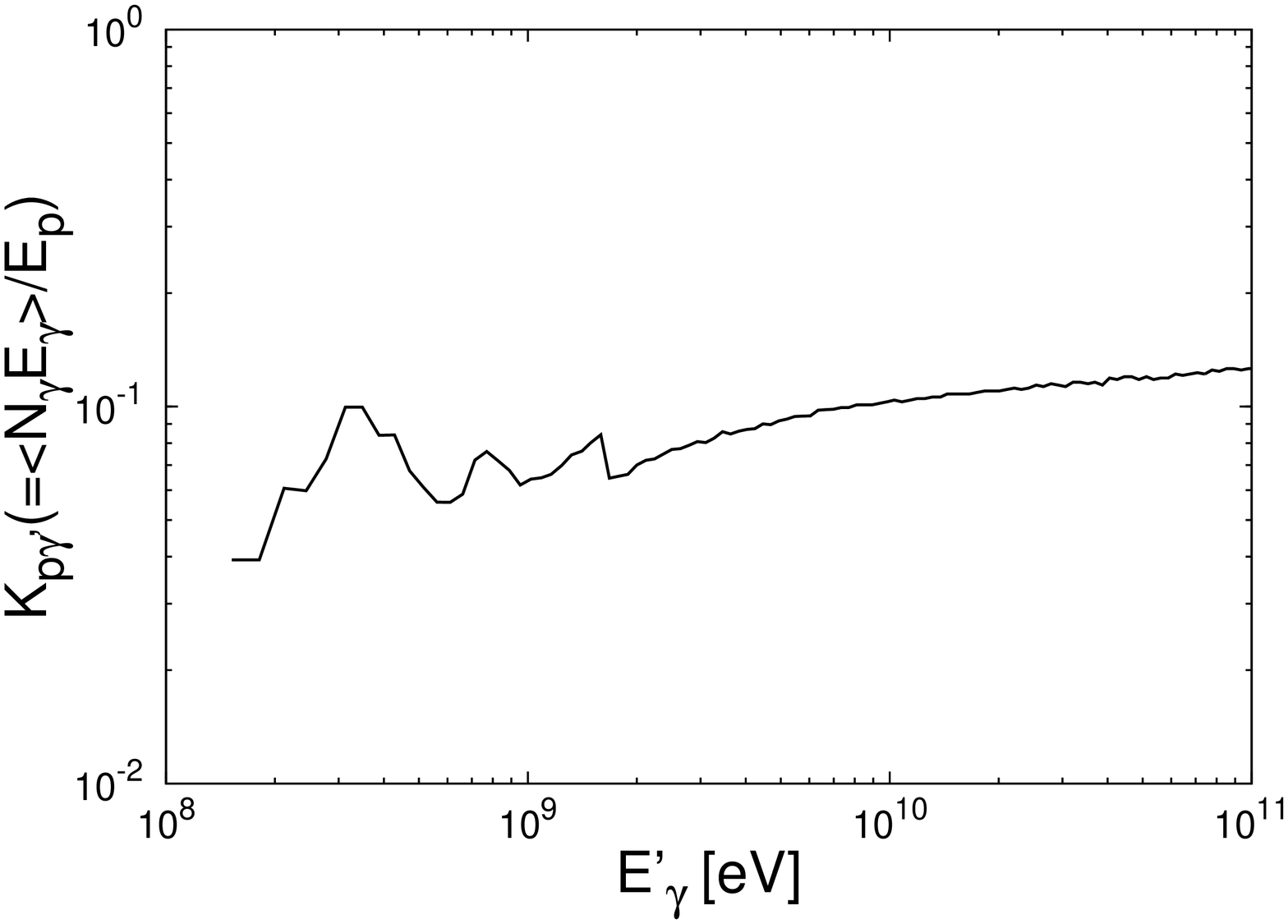}
\caption{The cross-section (left) and energy exchange per interaction (right) for $p\gamma$ interactions as a function of the photon energy in the proton's rest frame. These plots were obtained using the results from \cite{Mucke:1999yb}.}
\end{figure}

{\bf Photo-disintegration Interactions}-\\
This process has a cross-section containing a resonance (the Giant Dipole Resonance)
whose height can be approximately described by 
$\sigma_{A\gamma}^{\rm th.}\approx \sigma_{\rm mb}A\nonumber$. A parameterisation
of the measured form for certain species is provided by \cite{Khan:2004nd}.

The energy exhange to photons per interaction is
\begin{eqnarray}
K_{A\gamma}=\frac{\langle N_{\gamma}E_{\gamma}\rangle}{E_{A}} \approx \frac{10^{-3}}{A}.
\end{eqnarray}
where $\langle N_{\gamma}\rangle $ is the mean number of photons liberated in each interaction, $\langle E_{\gamma} \rangle$ is
the mean energy of each photon produced in the lab frame, and $E_{A}$ is the nucleus's energy in the lab frame.

\begin{figure}[h!]
\centering\leavevmode
\includegraphics[width=0.33\linewidth,angle=-90]{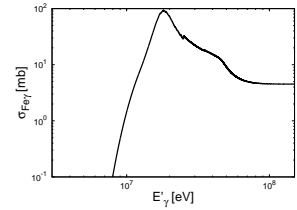}
\caption{The cross-section for Fe$\gamma$ interactions as a function of the photon energy in the nuclei's result frame. This plot was obtained using the results from \cite{Khan:2004nd}.}
\label{Fegamma_inelasticity}
\end{figure}

\begin{widetext}

Using the cross-sections given above, the interaction lengths for each process may be obtained 
\begin{eqnarray}
%\hspace{-1cm}
t_{\rm {\rm CR}\gamma}^{\rm int}(E_{\rm CR}) &=& \left(\frac{(m_{\rm CR}c^{2})^{2}}{2E_{\rm CR}^{2}}\int_{0}^{\infty}\frac{1}{\epsilon_{\gamma}^{2}}\frac{dn_{\gamma}}{d\epsilon_{\gamma}}d\epsilon_{\gamma}\int_{0}^{2E_{\rm CR}\epsilon_{\gamma}/(m_{\rm CR}c^{2})}\sigma_{{\rm CR}\gamma}(\epsilon'_{\gamma})\epsilon'_{\gamma}d\epsilon'_{\gamma}\right)^{-1}\nonumber 
\end{eqnarray}
%\end{widetext}

Combining the energy exchange per interaction with the interaction time, an expression
for the cooling time is obtained,
%%\begin{widetext}
\begin{eqnarray}
%\hspace{-1cm}
t_{\rm {\rm CR}\gamma}^{\rm cool} (E_{\rm CR})&=& \left(\frac{(m_{\rm CR}c^{2})^{2}}{2E_{\rm CR}^{2}}\int_{0}^{\infty}\frac{1}{\epsilon_{\gamma}^{2}}\frac{dn_{\gamma}}{d\epsilon_{\gamma}}d\epsilon_{\gamma}\int_{0}^{2E_{\rm CR}\epsilon_{\gamma}/(m_{\rm CR}c^{2})}K_{{\rm CR}\gamma}(\epsilon'_{\gamma})\sigma_{{\rm CR}\gamma}(\epsilon'_{\gamma})\epsilon'_{\gamma}d\epsilon'_{\gamma}\right)^{-1}\nonumber
\end{eqnarray}
\end{widetext}

\end{appendix}

\end{document}